\documentclass[preprint,12pt]{elsarticle}

\usepackage{amssymb}
\usepackage{amsmath}

\usepackage[version=4]{mhchem}
\usepackage{siunitx}
\usepackage{enumitem}
\usepackage{mathtools}
\usepackage{fancyvrb} 
\usepackage{pdfsync}
\usepackage{boldline}
\usepackage{longtable}
\usepackage{anysize}
\usepackage{cprotect}
\usepackage{graphicx}
\usepackage{epsfig}
\usepackage{verbatim} 
\usepackage{bm}
\usepackage{etoolbox}
\usepackage{multirow}
\usepackage{hyperref}
\usepackage[capitalise]{cleveref}

\crefname{enumi}{point}{points}
\usepackage{acro}
\acsetup{make-links = true}

\providecommand{\mytitile}{Parallelization of Gillespie algorithm based on binary words}

\providecommand{\bal}{\begin{align}}
\providecommand{\eal}{\end{align}}

\providecommand{\be}{\begin{equation}}
\providecommand{\ee}{\end{equation}}
\providecommand{\bsp}{\begin{split}}
\providecommand{\esp}{\end{split}}
\providecommand{\bea}{\begin{eqnarray}}
\providecommand{\eea}{\end{eqnarray}}
\providecommand{\beas}{\begin{eqnarray*}}
\providecommand{\eeas}{\end{eqnarray*}}

\providecommand{\bw}{\begin{widetext}}
\providecommand{\ew}{\end{widetext}}

\providecommand{\ra}{\rightarrow} 

\providecommand{\bn}{{\bm N}}
\providecommand{\bb}{{\bm b}}
\providecommand{\bc}{{\bf c}}
\providecommand{\bs}{{\bf s}}
\providecommand{\bma}{{\bm a}}
\providecommand{\nl}{N_{\rm L}}
\providecommand{\nr}{N_{\rm R}}
\providecommand{\na}{N_{\rm A}}

\providecommand{\bnl}{{\bm N}_{\rm L}}
\providecommand{\bnr}{{\bm N}_{\rm R}}
\providecommand{\bna}{{\bm N}_{\rm A}}

\providecommand{\cn}{{\mathcal N}}
\providecommand{\bcn}{{\bm {\mathcal N}}}
\providecommand{\ntot}{N_{\rm tot }}
\providecommand{\co}{{\mathcal O}}
\providecommand{\changer}{{\bm \xi}}
\providecommand{\comp}{{\bm \omega}}
\providecommand{\xor}{\,  \veebar \, }
\providecommand{\myland}{\,  {\&} \, }

\providecommand{\registered}{{\textsuperscript \textregistered} }
\providecommand{\bnlc}{{\bnl^\changer}}
\providecommand{\bnrc}{{\bnr^\changer}}
\providecommand{\bnac}{{\bna^\changer}}

\DeclareAcronym{CDF}{
	short = CDF, 
	long = cumulative distribution function
}
\DeclareAcronym{SSE}{
	short = SSE, 
	long = streaming \ac{SIMD} extensions
}
\DeclareAcronym{GPU}{
    short = GPU, 
    short-plural-form = {GPUs},
    long = graphics processing unit,
    long-plural-form = {graphics processing units}
    }
\DeclareAcronym{L}{
		short = L, 
		long = left handed,
}
\DeclareAcronym{D}{
		short = D, 
		long = right handed,
}
\DeclareAcronym{A}{
		short = A, 
		long = activator,
}
\DeclareAcronym{SIMD}{
		short = SIMD, 
		long = {single instruction, multiple data},
}
\DeclareAcronym{IEEE}{
		short = IEEE, 
		long = {Institute of Electrical and Electronics Engineers},
}

\journal{Journal of Computational Physics}

\begin{document}

\begin{frontmatter}


\title{\mytitile}

\author{Michele Castellana}
\affiliation{organization={Institut Curie, PSL Research University, CNRS UMR168},
            addressline={11 rue Pierre et Marie Curie}, 
            city={Paris},
            postcode={75005}, 
            country={France}}

\author{David Lacoste}
\affiliation{organization={Gulliver Laboratory, UMR CNRS 7083, PSL Research University, ESPCI},
            addressline={10 Rue Vauquelin}, 
            city={Paris},
            postcode={F-75231}, 
            country={France}}

\begin{abstract}
We present an improvement of the Gillespie Exact Stochastic Simulation Algorithm, which leverages a bitwise representation of variables to perform independent simulations in parallel. We show that the subsequent gain in computational yield  is significant, and it may allow to perform simulations of non-well mixed chemical systems. We illustrate this idea with  simulations of Frank model, originally introduced to explain the emergence of homochirality in prebiotic systems.
\end{abstract}




\end{frontmatter}

\renewenvironment{description}
{\list{}{\labelwidth0pt\itemindent-\leftmargin
    \parsep0pt\itemsep-10pt\let\makelabel\descriptionlabel}}
               {\endlist}

\setlength{\parskip}{5pt plus 0pt minus 0pt}

\section{Introduction}\label{section-intro}

Stochasticity plays a central role in biological systems, and for this reason the dynamics of cellular processes requires particular attention. The most common formal approach to model the dynamics of these processes uses a deterministic and continuous formalism, based on ordinary differential equations. While this representation is often adequate, there are also many cases in chemistry and biology where it fails, because the cellular process of interest involves a small number of molecules. This concerns, for instance, modeling of single molecular motors or of gene expression in single cells, which is typically controlled by a low number of mRNA molecules.

In all these cases, it is necessary to resort to a discrete, stochastic approach, based on the solution of the master equation via the Gillespie algorithm \cite{gillespie1976general}. This approach takes into account the fact that at the molecular level, species numbers change in discrete, integer amounts. In addition, changes in the number of molecules are assumed to result from the inherent random nature of microscopic molecular collisions; in this sense, the method  describes the molecular processes at work in the cellular space. However, the approach has a major limitation: when simulating large systems, it becomes computationally  expensive. Essentially, any method based on the simulation of a reaction event one at a time suffers from this problem.

Since the discovery of the algorithm in the seventies, many improvements have been made to address the issue. In one improvement of the original algorithm, species and reactions are introduced only when they are needed \cite{Lok2005}. In another one, called tau-leaping method, reactions are grouped together assuming their propensities do not change significantly in a certain time interval $\tau$, and then concentrations are jumped forward in time using a sum of independent Poisson variables with a mean proportional to that time \cite{Gillespie_2001}. If, in addition,  many events are assumed to contribute to this jump, one obtains from the tau-leaping formula the so-called chemical Langevin equation, which describes chemical systems with continuous random variables at the level of Gaussian fluctuations \cite{gillespie2007stochastic}. 
This approach is significantly faster than the Gillespie algorithm. For this reason, 
T. C. Elston introduced hybrid models which treat certain variables as discrete, by using an efficient implementation of the Gillespie  algorithm, while other variables are treated as continuous using the chemical Langevin equation \cite{adalsteinsson_2004}. In all these improvements, however, the system of interest is still assumed to be well mixed. In fact, according to Gillespie ``the problem of how best to simulate systems that are not well stirred [...] holds a great many challenges'' \cite{gillespie2007stochastic}. 
In this regard, in 2012 a new variant of the algorithm has been proposed, which achieves a significant acceleration by leveraging the power of parallel computing using \acp{GPU} \cite{Komarov_2012}. 
Here, we present a different strategy of parallelization of the Gillespie algorithm, which does not require a complex implementation on \acp{GPU}, but which is instead based on a simpler idea, namely a bitwise representation of variables. 

To provide a proof of concept of the efficiency of the method, we use a classic model which has been originally introduced to explain the emergence of homochirality in prebiotic systems---the Frank model \cite{frank1953on}. In  \cref{section-frank} we present the model, and in \cref{standard-gillespie} we discuss the classic Gillespie algorithm  to simulate it. In \cref{bitwise-gillespie} we detail the bitwise Gillespie algorithm, show dynamical trajectories obtained with it, and in  \cref{efficiency-gain} we study in  the efficiency gain with respect to the classical algorithm. Finally, \cref{section-discussion} is dedicated to the discussion and to the future directions of this work. Throughout the paper, we will refer to our implementation of the bitwise algorithm in our \verb|C++| \cite{noauthor2020iso} code available on \cprotect{\href{https://github.com/mcastel1/gillespie}}{Github}, by including links to the code which implements the concepts presented in the manuscript. 

\section{Frank's model}\label{section-frank}

The Frank model involves three molecular species \cite{frank1953on}: a \ac{L}, a \ac{D} and an \ac{A}, whose numbers will be denoted by $\nl$, $\nr$ and $\na$, respectively. The \ac{L} and \ac{D} species represent two enantiomeric forms of the same molecules, while species \ac{A} is achiral. The state of the system is given by the three occupancies $\cn \equiv \{ \nl , \nr, \na \}$.  These species  undergo the  following reactions:

\begin{align}
\label{eq-frank-1}
\rm A + \rm L 
\xrightarrow{k_{\rm A}} & 2\, \rm L, \\ 
\label{eq-frank-2}
\rm A + \rm R \xrightarrow{k_{\rm A}} & 2\,  \rm R, \\ 
\label{eq-frank-3}
\rm L + \rm R \xrightarrow{k_{\rm I}} & 2\, \rm A,
\end{align}
where $k_{\rm A}$ and $k_{\rm I}$ denote the transition rates. The first two reactions describe the autocatalytic amplification of enantiomer \ac{L} and \ac{D}, while the last one represents a chiral inhibition reaction, in which the two enantiomeric species \ac{L} and \ac{D} combine.
As a proof of concept of our algorithm, and for illustrational purposes only, we will now consider a modified version of Frank's model, in which we have, in addition to \cref{eq-frank-1,eq-frank-2,eq-frank-3,eq-frank-4,eq-frank-5,eq-frank-6}, three \textit{ghost} reactions

\begin{align}
	\label{eq-frank-4}
	\rm L + \rm L \xrightarrow{} & 2\, \rm L, \\ 
	\label{eq-frank-5}
	\rm R + \rm R \xrightarrow{} & 2\,  \rm R, \\ 
	\label{eq-frank-6}
	\rm A + \rm A \xrightarrow{} & 2\, \rm A. 
\end{align}

Despite the fact that they are immaterial, \cref{eq-frank-4,eq-frank-5,eq-frank-6} will prove useful in implementing the bitwise strategy for the model dynamics---see \cref{gillespie-nobit-3} below.
In addition, we  set $k_{\rm A} = k_{\rm I} = 1$ in \cref{eq-frank-1,eq-frank-2,eq-frank-3}, in such a way that all propensity functions are integers.

 In what follows, the reactions will be labelled by index $1 \leq r\leq 6$, and we will denote by 
\be\label{equation-as}
\begin{aligned}
	a_1(\cn) \equiv &\na \nl,\\
	a_2(\cn) \equiv &\na \nr, \\
	a_3(\cn) \equiv &\nl \nr, \\
a_4(\cn) \equiv &\nl(\nl-1)/2, \\
a_5(\cn) \equiv &\nr(\nr-1)/2, \\
a_6(\cn) \equiv &\na(\na-1)/2 .
\end{aligned}
\ee
the \textit{propensity functions} \citep{gillespie2007stochastic}, in which we indicated explicitly the dependency on the system's state. 

Finally, we observe that  \cref{eq-frank-1,eq-frank-2,eq-frank-3,eq-frank-4,eq-frank-5,eq-frank-6}  conserve the total number of particles $\ntot \equiv \nl + \nr + \na$; this property  will prove useful in the bitwise implementation of the Gillespie algorithm, see \cref{point-bit-iteration-1} of \cref{bitwise-gillespie}.

\section{Gillespie algorithm}\label{standard-gillespie}

The idea behind Gillespie algorithm is a simple, exact equation for the probability of the time to the next reaction and the type of such reaction, which results in update rule which we will describe below  \citep{gillespie2007stochastic}. For the sake of clarity, we will tailor the presentation of the update rule so as to establish a clear analogy with the parallel update rule which will be presented in \cref{bitwise-gillespie}. 

The update rule is given by the following steps: 
\begin{enumerate}
\item  \label{gillespie-nobit-1} Given the state  $\cn$ of the system at time $t$.

\item  \label{gillespie-nobit-4}  Draw the time lapse  $\tau$ for the next reaction from the distribution 
\be\label{equation-tau}
p(\tau) = {Z(\cn)} e^{-Z(\cn)\tau}.
\ee
\item  \label{gillespie-nobit-2}  Evaluate all the propensity functions $a_j(\cn)$, given by \cref{equation-as}.
\item \label{gillespie-nobit-3}  Evaluate the normalization
\be\label{equation-z}
\begin{aligned}
Z(\cn) \equiv & \sum_j a_j(\cn)\\
  = & \frac{1}{2} \Bigg[ \sum_{i \neq j} \cn_i \cn_j + \sum_i \cn_i (\cn_i-1)\Bigg]  \\
= & \frac{\ntot(\ntot-1)}{2}. 
\end{aligned}
\ee
Note that in the second line of \cref{equation-z} we used the ghost reactions, \cref{eq-frank-4,eq-frank-5,eq-frank-6}, which allow us to rewrite $Z$ in terms of the total number of particles $\ntot$, by adding to the off-diagonal terms  $ \sum_{i \neq j} \cn_i \cn_j$ the diagonal terms $ \sum_i \cn_i (\cn_i-1)$.  Given that  the total number of particles $\ntot$ is conserved, the normalization of the modified model is also conserved. 

\item  \label{gillespie-nobit-5} Draw the index $r$ of the next reaction according to the \ac{CDF} 
\be 
F_r = \frac{1}{Z(\cn)}\sum_{p=1}^r a_p(\cn).
\ee

This can be achieved as follows:

\begin{enumerate}[label=\roman{*}), ref=(\roman{*})]
	\item \label{iteration-nobits-F-1} Draw a random integer $n$ uniformly distributed in $[0, Z-1]$,
	\item \label{iteration-nobits-F-2} Set  $r = 0$	and consider an integer  $m=0$,
	\item \label{iteration-nobits-F-3} Compare $m$ with $n$,
	\item \label{iteration-nobits-F-4} If $m \leq n$, set 
	\be
	m \ra m + a_r(\cn), \,  r \ra r+1,
	\ee
	  and go to \cref{iteration-nobits-F-3}. Stop otherwise. 
\end{enumerate}

The resulting $r$ is the index of the reaction to effect, see \cref{figure-F}.\\

\item \label{gillespie-nobit-6}  Effect the reaction:

\begin{itemize}
\item  Update the time 
\be\label{eq-update-time}
t \ra t + \tau,
\ee
\item  Update the occupancies according to \cref{eq-frank-1,eq-frank-2,eq-frank-3}: 
\be\label{eq-update-occupancies}
\cn \ra \cn + \nu_r,
\ee
where, according to \cref{eq-frank-1,eq-frank-2,eq-frank-3},  the state-change vector $\bm \nu$ \citep{gillespie2007stochastic} is given by 
\begin{equation}\label{eq_nu}
	\begin{aligned}
	\nu_1 = & \{1, 	0, -1\}, \\
	\nu_2 = & \{0, 	1, -1\}, \\
	\nu_3 = & \{-1, 	-1, 2\}.
	\end{aligned}
\end{equation}
and the other components of $\bm \nu$ are identically zero.

\item Update the propensity functions according to \cref{equation-a-nobit-1,equation-a-nobit-2,equation-a-nobit-3}, see \cref{app-1} for details.

\end{itemize}

\item\label{gillespie-nobit-7} Go to \cref{gillespie-nobit-1}.
\end{enumerate}

In what follows, we will propose a method, inspired by Monte Carlo simulations of disordered systems \cite{palassini1999universal},  to increase the computational yield of this algorithm by leveraging the discrete nature of its degrees of freedom, and the binary nature of data  stored in binary words.

\section{Bitwise Gillespie algorithm}\label{bitwise-gillespie}

Given an occupancy number $N$ in the standard Gillespie algorithm, such as $\nl$, $\nr$ and $\na$, we write it in base $2$ as
\be \label{eq3} 
N = \sum_{i=0}^{N_2} b_i 2^i,
\ee
where $b_i = \{0,1\}$ are the components of $N$ in base two, and $N_2$ the minimal  number of bits necessary to write $N$ in such base. 

We will now leverage the parallel structure in which the $\bb$s are stored in a computer, by considering $M$ copies of the system \citep{palassini1999universal,freund1988multispin,bernaschi2024quisg}. As a result, $N$ will now represent a collection of $M$ integers
\be\label{eq1}
\bn = \{ N^1, \cdots, N^M \},  
\ee
and each $\bb_i$  a set of $M$ boolean variables, or a \textit{binary word} 
\be\label{eq2}
\bb_i = \{ b_i^1, \cdots, b_i^M \},
\ee 
where each $b_i^j $ is equal to either zero or one, and in what follows, binary words will be written in bold face.   The occupancy number $\bn$ is then given by the generalization of \cref{eq3}:
\be  \label{eq4}  
\bn = \sum_{i=0}^{\bn_2-1} \bb_i 2^i,
\ee
which must be read in a bitwise manner as 
\be \label{eq6}
N^j = \sum_{i=0}^{\bn_2-1} b_i^j 2^i.
\ee

For instance, if $M = 32$ or $64$, we can write $\bb_i$ as a  \verb|C++| \cite{noauthor2020iso} $64$-bit type  \verb+unsigned int+ or \verb+unsigned long int+, respectively \citep{stroustrup1995the}. In the \verb|C++| code relative to this work, $\bb_i$ and $\bn$ are given by \cprotect{\href{https://github.com/mcastel1/gillespie/blob/master/include/bits.hpp}}{\verb|Bits|} and an \cprotect{\href{https://github.com/mcastel1/gillespie/blob/master/include/int.hpp}}{\verb|Int|} class objects, respectively.

We will now implement the Gillespie iteration by using the bitwise implementation. 
To achieve this, we  will assume that the $M$ system copies share the \textit{same} total number of particles $\ntot$. Given that the dynamics conserves $\ntot$, see \cref{standard-gillespie}, setting each copy with an independent, random initial condition and running in parallel the dynamics of the $M$ copies is tantamount to increasing the number of temporal samples resulting from the simulation. \\

We will now implement   \cref{gillespie-nobit-1,gillespie-nobit-2,gillespie-nobit-3,gillespie-nobit-4,gillespie-nobit-5,gillespie-nobit-6,gillespie-nobit-7} of the parallel Gillespie. While \cref{gillespie-nobit-1,gillespie-nobit-2,gillespie-nobit-3,gillespie-nobit-4} are identical to their serial version presented above, \cref{gillespie-nobit-5,gillespie-nobit-6} require particular care to be implemented in a parallel manner, as discussed in the following. 

\begin{figure}
	\centering
	\includegraphics[width=0.6\linewidth]{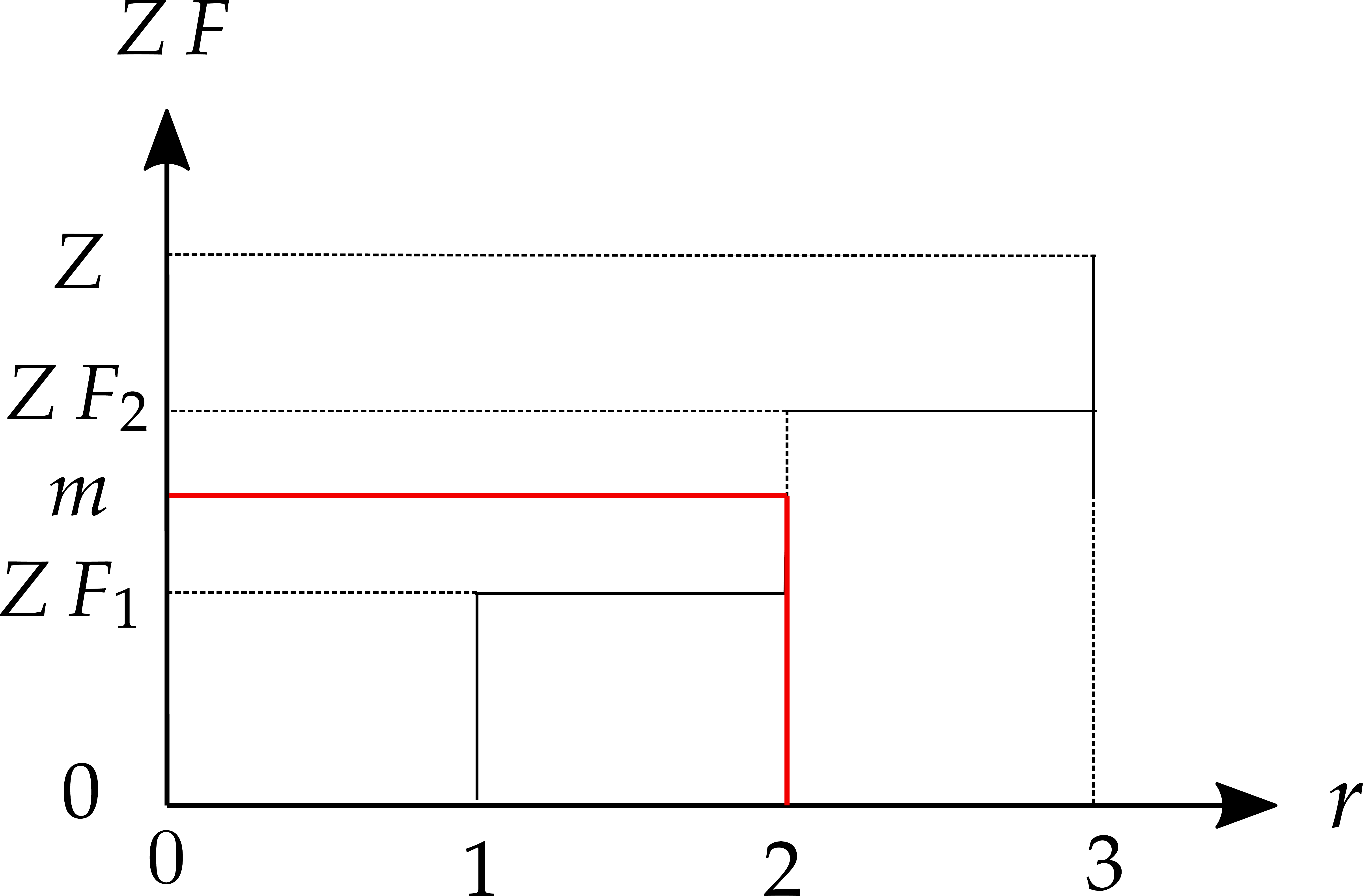}
	\caption{Drawing the index of the reaction. A random integer $m$ is drawn in the interval $[0, Z-1]$, where $Z$ is an integer. Starting with $r=0$, one runs through all reactions, and stops when $Z F_r$ is larger than $m$. The resulting value $r=2$ is the reaction drawn  (red lines). }
	\label{figure-F}
\end{figure}

In the serial case the reaction is drawn by running through the values of the reaction index $r$, and stopping when $Z F_r$ reaches the threshold $m$, see \cref{figure-F}. In lieu of this standard procedure, in the parallel iteration of Gillespie algorithm a different strategy must be employed. In fact, for a given $m$, among the $M$ copies of the system which we consider in parallel, different copies may reach the threshold at different values of $r$, making the serial, standard approach above inapplicable. Instead, we will draw the reaction in parallel by  merging \cref{gillespie-nobit-5,gillespie-nobit-6} of the Gillespie iteration above into a single step, and by drawing and effecting the reaction by means of a \textit{changer word}---see below.  

We will rewrite \cref{gillespie-nobit-5,gillespie-nobit-6} as follows: 

\begin{enumerate}[label=\alph*)]
	\item \label{point-bit-iteration-1} Draw a random integer $n$ uniformly distributed in $[0, Z-1]$, and consider  $M$ identical copies of it 
	\be
	{\bm n} \equiv \{ n , \cdots, n \}
	\ee 
	as in \cref{eq1}. Given that $Z$ is independent of time, see \cref{equation-z}, and identical for all copies of the system, this allows for drawing the \textit{same} random number for all copies---see below. 
	\item Set $r=0 $, consider  the integer ${\bm m} = \{ 0, \cdots, 0 \}$ and  the \textit{comparator} word 
	\be
	\comp \equiv {\bf 1} \equiv \{1, \cdots, 1\},
	\ee 
	which wil be used to compare integer numbers in a bitwise manner. 
	\item \label{iteration-bits-F-3} Compare $\bm m$ with $\bm n$:

	\begin{itemize}
		\item Compute the  word 
		\be
		 {\bm m} \leq {\bm n}, 
		\ee 
		defined by a bit-wise comparison between $\bm m$ and $\bm n$:

		\be
		[{\bm m} \leq {\bm n}]^i = \left\{
		\begin{array}{ll}
			1 & \textrm{ if } m^i \leq n^i \\
			0 & \textrm{ otherwise.},
			\end{array}\right.
		\ee
		\item Compute the changer word: 
		\be\label{equation-changer}
		\changer \equiv \comp \xor  [{\bm m} \leq {\bm n}],
		\ee
		where $\xor$ denotes the bitwise XOR between the two words \cite{bochenski1959precis}.

		The word $\changer$ detects if, by running through the reactions $r \ra r+1$, the \ac{CDF} is larger than  the threshold  $\bm n$, along the lines of the procedure shown in  \cref{figure-F}. If this is the case for the $i$th copy of the system, then $\xi^i = 1$, and such copy will  undergo the $r$th reaction, while $\xi^i = 0$ otherwise, see \cref{point-update-occupencies}.

	\end{itemize}

	\item Update $\comp$ and $\bm m$ for the the next iteration: 
	Set 
	\be
	\comp = [{\bm m} \leq {\bm n}]
	\ee 
	and 
	\be
	{\bm m} \ra {\bm m} + {\bm a}_j(\bcn).
	\ee

	\item Update the occupancies: \label{point-update-occupencies}
		
		The  definition  of the changer word in \cref{equation-changer} allows us to perform the updates of the occupancies in \cref{eq-update-occupancies} in a bitwise manner. In fact, if the $i$th copy of the system undergoes the $r$th reaction, then $\changer^i$ is equal to one, and zero otherwise. It follows that, from \cref{eq-update-occupancies}, we  set 
		\be\label{equation-v}
		\bcn \ra  \bcn + [\changer \myland v_r],
		\ee
		where $\myland$ denotes the bitwise AND between the two words \cite{bochenski1959precis}, and we perform in parallel the occupancy update for all copies.

		\item Update the propensity functions. 
		Given an integer $\bn$ represented as in \cref{eq1,eq2,eq4,eq6} and a binary word, e.g., $\changer$, we set
		\be
		\bn \myland \changer \equiv  \sum_{i=0}^{\bn_2-1} [\bb_i \myland \changer] \, 2^i,
		\ee
		and define the changed occupancies as follows: 
		\begin{equation}
			\begin{aligned}
				\bnlc = &\bnl \myland \changer,\\
				\bnrc = &\bnr \myland \changer,\\
				\bnac = &\bna \myland \changer.
			\end{aligned}
		\end{equation}

		Proceeding along the lines of \cref{point-update-occupencies}, the above definition of $\bnlc$, $\bnrc$ and $\bnac$ allows us to perform the updates of the occupancies and propensity functions in \cref{equation-a-1,equation-a-2,equation-a-3} in a bitwise manner. In fact, given that the multiplication operation in \cref{equation-a-nobit-1,equation-a-nobit-2,equation-a-nobit-3} corresponds to a logical $\myland$ between binary words, the update for the propensity functions is given by \cref{equation-a-1,equation-a-2,equation-a-3}, see \cref{app-2}.

\item Set $r \ra r+1$ and go to \cref{iteration-bits-F-3}

\end{enumerate}

Finally, one updates the time with \cref{eq-update-time} and goes back to \cref{gillespie-nobit-1}.  

Note that in the algorithm above, we used the same random number for all copies of the system. On the one hand, if each of the $M$ copies of the system is initialized with an independent initial condition, the $M$ dynamics which evolve in parallel can be considered as independent, and they constitute an increase of a factor $M$ of the number of samples \cite{palassini1999universal}. 
On the other hand, given that the random number constitutes the step of the Gillespie iteration which is computationally  costly, this parallel procedure will allow us to increase the computational  yield with respect to the serial case---see below.

\subsection{Arithmetic}\label{arithmetic}

Given the bitwise representation above, we will now show how to make arithmetic operations in parallel with the occupancy numbers $\bn$ \citep{palassini1999universal}. We will now focus on the arithmetic operations which appear in the Gillespie iteration, \cref{equation-v,equation-a-1,equation-a-2,equation-a-3}. 

\subsubsection{Sum and substraction} 

\Cref{equation-v,equation-a-1,equation-a-2,equation-a-3}  involve sums between integers, which we  compute in a bitwise manner as follows. 

Given two occupancy numbers $\bn_1$, $\bn_2$ represented as in \cref{eq4}, their sum $\bn_+ \equiv \bn_1 +\bn_2$ can be  written in terms of binary coefficients, which we will denote by $\bb_+$, as per \cref{eq4}. The coefficients $\bb_+$ are obtained from $\bb_1$ and $\bb_2$ by performing bitwise operations only, where in each of these operations the $M$ copies of the system are treated at once, see  \cprotect{\href{https://github.com/mcastel1/gillespie/blob/master/include/int.hpp}}{\verb|Int::AddTo|}. We perform the sum with a standard, carry-based algorithm where one runs over the entries of $\bb_1$ and $\bb_2$ and, for each of them, compute the result of the sum and the carry. The carry is thus propagated bitwise through the  entries of the sum, and the result stored in $\bb_+$. 

To illustrate this process, consider the following pseudocode, for which we assume that $\bb_1$ and $\bb_2$ contain the same number of entries, $N_2$,  see \cprotect{\href{https://github.com/mcastel1/gillespie/blob/master/include/int.hpp}}{\verb|Int::AddTo|}: 

\begin{itemize}
\item Set the carry $\bc = \bf 0$
\item For $i=0, \cdots, N_2-1$ 
\begin{itemize}
\item Update the partial sum $\bs = {\bb_1}_i \, \& \, {\bb_2}_i $
\item Compute the carry of the partial sum $\bc = [{\bb_2}_i \, \& \, ({\bb_1}_i \, \| \, \bc )] \, \| \, ({\bb_1}_i \, \& \, \bc ) $
\item Update ${\bb_+}_i = \bs$
\end{itemize}
\item Return the result $\bn_+$ and the carry $\bc$.
\end{itemize}

The logical operations \verb+&+ and \verb+|+ above are carried on the $M$ entries of their arguments at once: for example
\begin{equation}
\begin{aligned}
	{\bb_1}_i = & \{ 0 \, 1 \, 0\, 0 \cdots 1 \, 1\, 1 \},\\
	{\bb_2}_i = & \{ 1 \, 1 \, 1\, 0 \cdots 1 \, 0\, 0 \},\\
	{\bb_1}_i \, \& \, {\bb_2}_i = & \{ 0 \, 1 \, 0\, 0 \cdots 1 \, 0\, 0 \}.
\end{aligned}
\end{equation}

\begin{figure}
	\centering
	\includegraphics[width=0.6\linewidth]{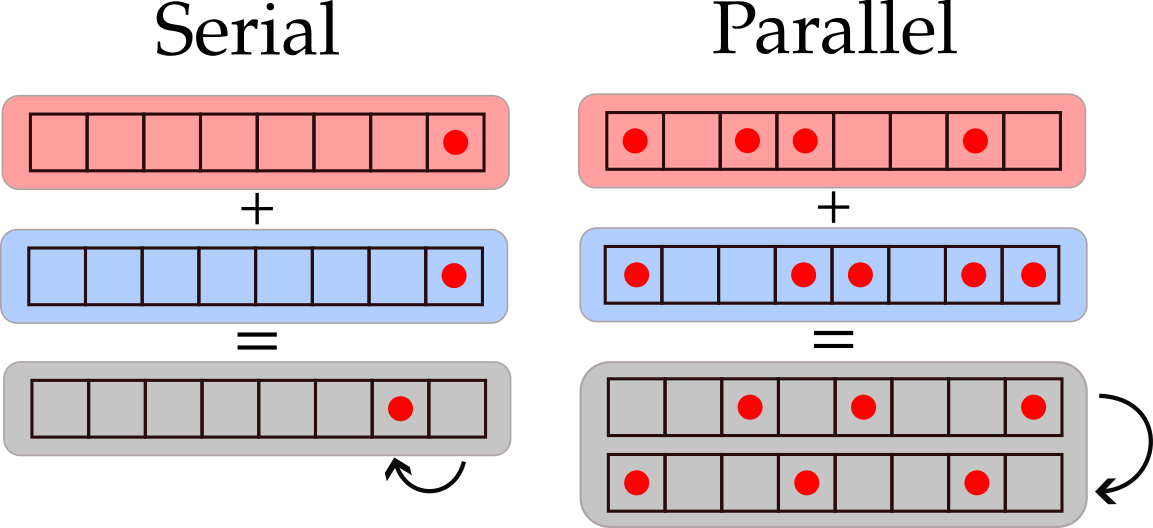}
	\caption{A sum performed with logical operations. 
	Left: sum performed in a standard, serial way. 
	The augend (the number to which the addend is summed), the addend (the number summed to the augend), and the sum are represented in the computer as a collection of eight bits, where each bit is stored in a  square. Each bit can be either zero or one, and it corresponds to the presence or the absence of voltage in the hardware chip (red dot). Bits are read from right to left, and they correspond to the coefficients of the powers of $2^0,2^1, \cdots$, respectively.
	The augend (red rectangle) and the addend (blue) are both equal to one, and the sum (gray) is equal to two.  The black arrow denotes the direction in which the carry of the sum propagates. 
	Right: sum performed in a parallel way. The augend (red) and addend (blue) represent a collection of eight integers (replicas, each stored in a column), which are summed in parallel. Each of these can be either zero (no red dot) or one (red dot).  The sum is shown in gray, and its first and second row correspond to the coefficient of $2^0$ and $2^1$, respectively. The black arrow is the direction in which the carry of the sum propagates. 
	}
	\label{figure-parallel}
\end{figure}

Given that each $\bb$ is stored as a collection of stacked bits in the computer memory, the $\&$ operation above is performed \textit{in parallel} across all such bits. This implies a considerable computational gain with respect to the standard Gillespie algorithm, where arithmetic operations are carried on serially.  \Cref{figure-parallel} shows this gain explicitly in a minimal example, where two, one-bit integers are added:  In the serial case (left), we perform one sum, while in the  parallel (right) case, we carry on $M$ sums in parallel, with a computational cost comparable to the serial case. This increases the yield by a factor comparable to $M$.

Proceeding along the same lines, one can perform a bitwise substation between two integers which appear in \cref{equation-v,equation-a-1,equation-a-2,equation-a-3}, i.e.,  $\bn_- = \bn_1 - \bn_2$ for which $N_1 ^j \geq N_2^j$ for all $j$s. Among the multiple available  subtraction algorithms \citep{surhone2010twos}, we found that the one more suited to this problem is the classical, borrow-based subtraction, which is the analog of the carry-based algorithm used above for the sums, see \cprotect{\href{https://github.com/mcastel1/gillespie/blob/master/include/int.hpp}}{\verb|Int::SubstractTo|}.  

\subsubsection{Multiplication}

While the bitwise sum between two numbers of $N_2$ bits each involves $\co(N_2)$ operations, a multiplication  between these two numbers would involve $\co(N_2^2)$ operations, and it would thus be computationally costly for large occupancies. It is for this reason that, in \cref{equation-a-1,equation-a-2,equation-a-3}, we rewrote the update rules by avoiding, as much as possible, integer multiplications.  In particular, \cref{equation-a-3} involves a multiplication by two only, which can be easily handled by shifting the binary word of the multiplicand by one bit \citep{surhone2010twos}, see \cprotect{\href{https://github.com/mcastel1/gillespie/blob/master/include/int.hpp}}{\verb|Int::MultiplyByTwoTo|}.

\section{Efficiency gain}

\label{efficiency-gain}

In what follows, we will study quantitatively the efficiency gain of the parallel Gillespie algorithm presented above. We stress the fact that this gain is not a gain in speed, but in the data yield of the simulation. Namely,  for a given amount of time and computational resources, the parallel Gillespie algorithm allows to increase the amount of sampled configurations by a factor, which we will denote by $Q$, with respect to the serial algorithm. 

\begin{figure}
	\centering
	\includegraphics[width=0.6\linewidth]{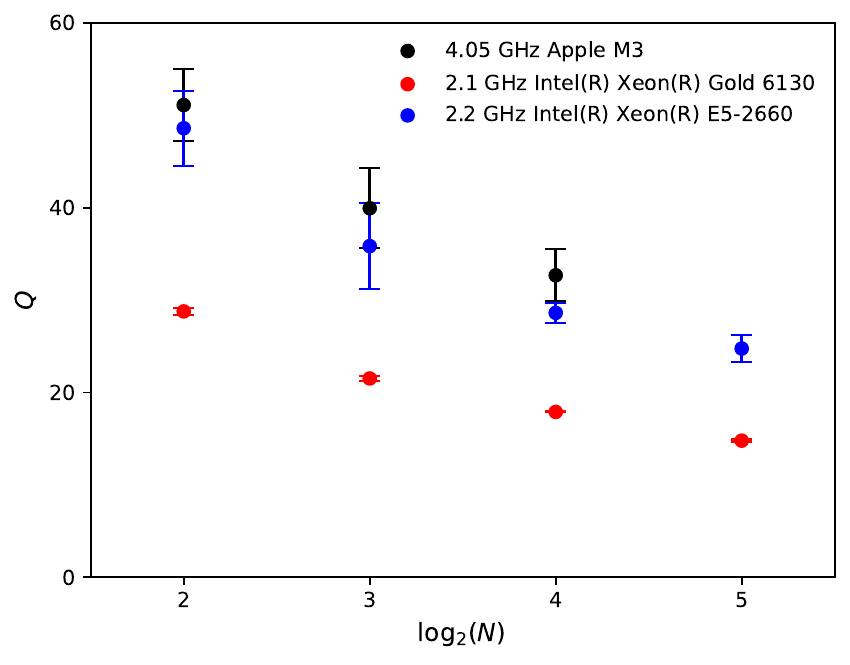}
	\caption{Computational gain $Q$ of the parallel (bitwise) Gillespie algorithm with respect to the serial (standard) algorithm, as a function of the total number of particles $N$ in the system, for $T = 10^5$ iterations and $M=64$ copies of the system (bits), on different architectures. The data has been obtained by making multiple runs of the Gillespie dynamics, and computing the average gain (dots) and its standard deviation (error bars) across the runs. }
	\label{figure-gain}
\end{figure}

Specifically, the gain $Q$ is computed by running the standard Gillespie algorithm $M$ times and the serial Gillespie algorithm once, simulating $M$ copies in parallel. Both algorithms perform the same number, $T$, of steps. Then 

\be
Q = \frac{\textrm{runtime of serial algorithm}}{\textrm{runtime of parallel algorithm}}, 
\ee
and the values of $Q$ are shown in \cref{figure-gain}. Some representative values   of $Q$ are shown in \cref{figure-gain}, which shows that the parallel algorithm yields a $\sim 15/60$-fold increase in performance for the values of $N$ which we considered.

\begin{figure}
	\centering
	\includegraphics[width=0.6\linewidth]{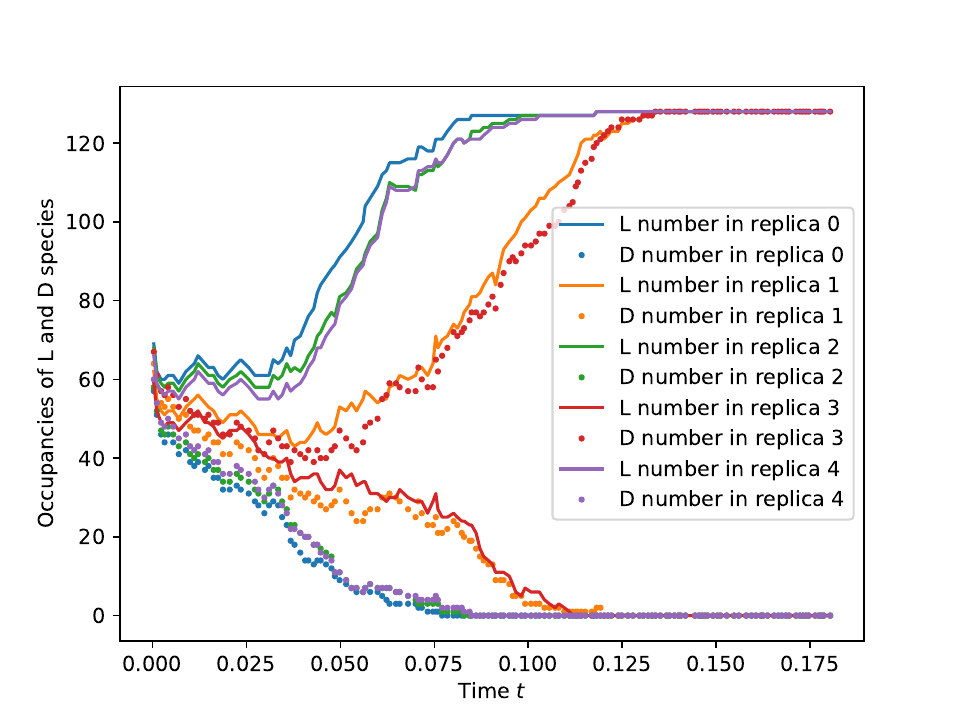}
	\caption{Simulation of the Frank model using the the parallel  Gillespie algorithm. 
	The solid (resp. dotted) curves represent the number of molecules of \ac{L} (\ac{D}) molecules 
	as a function of time for five among the 64 replicas simulated by the parallel algorithm.}
	\label{figure-frank}
\end{figure}

\section{Discussion}
\label{section-discussion}

We presented a novel method to increase the computational yield of numerical simulations of chemical-reaction networks with Gillespie algorithm \cite{gillespie1976general}, which leverages the binary fashion in which data is stored in binary words. This method is inspired, among others, from Monte Carlo  simulations of disordered systems, where the values of  Ising spins and  spin-spin couplings are stored in binary words \cite{palassini1999universal}.

In order to provide a proof of concept of the method, we considered  Frank model \cite{frank1953on} for homochirality---a minimal model where three chemical species, a left-handed, a right-handed and an activator, interact with each other. 
We let $M$ copies of the system evolve in parallel, by storing their occupancy numbers in a bitwise manner, and  by using the same random number for all copies. As shown in \cref{figure-gain}, given that the random-number generation is the bottleneck in the simulation, this yields a significant gain---comparable to $M$---in the simulation data yield. While  \cref{figure-gain} shows that this gain depends on the type of microprocessor, for  all  microprocessors which we tested, the parallel algorithm yields a significant  gain with respect to the serial one. 

The logic behind the increase in data yield is depicted in \cref{figure-parallel}, which shows that we can perform $M$ sums simultaneously  with a  computational effort similar to the one required for a single sum. The computational gain is comparable, but not equal to $M$. In fact, when parallelizing the operations as in the right panel of \cref{figure-parallel}, some efficiency is lost. This comes down to the fact, in the parallel approach, carry  propagation---represented  by arrows in \cref{figure-parallel}---is slower than in the serial case. In fact, the serial arithmetic benefits from  specific \textit{hardware} designs  which cannot be reproduced, in the present framework, in the parallel approach. A typical example of such  hardware structures is the carry-lookahead adder, which reduces the time needed to compute carry bits \cite{hurst1995contemporary}. 

A consequence of this slower carry propagation in the parallel algorithm is that the gain in computational yield is reduced when the total number of particles in the model increases, as shown in \cref{figure-gain}. As a result, our algorithm is mostly efficient in the regime of small number of particles. Fortunately, this is precisely the regime where the Gillespie algorithm is needed, because fluctuations are strong, and the approach based on the chemical Langevin equation is not appropriate to describe these fluctuations. 

Our parallel algorithm may prove useful, for example, in sampling rare events, by generating a larger number of samples with less computational effort 
with respect to the serial dynamics.
This is shown in \cref{figure-frank}, which depicts multiple dynamical trajectories for the numbers of \ac{L} and \ac{D} species 
in the Frank model for five copies among the 64 generated by the bitwise Gillespie algorithm. The initial distribution of the chemical species is chosen 
randomly according to a multinomial distribution centered around the racemic state. The figure shows that different initial conditions 
of the system lead to a full homochiral state which may be \ac{L} or \ac{D}, even though the same random numbers are used 
to generate the stochastic trajectories. This application illustrates the potential of the bitwise algorithm to sample a large set 
of initial conditions, which is particularly interesting when the system can transition between different 
steady-states, which happens in the Frank model.  One potential application of our algorithm could be the simulation of  Frank model 
of Ref. \cite{Hochberg_2023}, in which it was found that large simulation datasets were needed to get a proper confidence interval
to tell apart  stochastic and deterministic outcomes.
In addition to the Frank model, in research on the origin of life one is typically interested in studying a sequence of chemical
 and evolutionary transitions. These may be rare at first, but they can eventually occur due to a massive exploration of the chemical space of  species and reactions \cite{baum_2023}. 
In practice, this exploration of the chemical space is facilitated in large and non-well mixed systems,  because such systems 
often behave as an ensemble of independent smaller reactors or compartments, where each compartment samples a slightly 
different choices of initial composition or external conditions.
This mechanism is analogous to the bitwise algorithm, except that in the algorithm the evolution in each compartment 
is strictly independent, while this is not exactly true for prebiotic systems.

In future investigations, the performance of the parallel algorithm can be further increased by leveraging \ac{SIMD} extensions and, in particular,
 \ac{SSE}. In fact, \ac{SSE}s are special, large registers which contain  $M=128, 256$ and $512$ bits. The first, $128$-bit XMM \ac{SSE} was  introduced 
 in the Intel\registered Pentium\registered III microprocessors \citep{diefendorff1999pentium}. Later on, the   $256$-bit \ac{SSE}2 instructions were 
 introduced with the Intel\registered Pentium\registered IV \cite{intel2005pentium}. Finally, Intel\registered Xeon Phi x200 introduced  $512$-bit AVX
  registers  \citep{reinders2017intel}. We expect the gain of \cref{figure-gain}, obtained for $64$-bit registers, to be scalable and proportional to the register size. 

Another subject of future studies will be generalizing our approach to arbitrary values of the kinetic constants $k_{\rm A}$ and $k_{\rm I}$. Being physical quantities, 
these constants, as well as the propensity functions $a_1 = k_{\rm A} \na \nl$, $a_2 = k_{\rm A} \na \nr$, $\cdots$, will be  floating-point numbers which may be represented,
 for instance, with the \ac{IEEE} 754 standard, where a double-precision floating-point number is given by a sequence of $64$ bits---one for  the sign, eleven  for the 
 exponent, and fifty-three for the mantissa \cite{kahan1997lecture}. 
This representation is reminiscent of the binary format \eqref{eq3} for integer numbers, and it can thus be parallelized across multiple replicas along the lines 
of \cref{eq4}. 
If such reaction constants are incorporated in the dynamics, the update of the propensity functions will require  multiplications between floating-point numbers.
 These multiplications can be performed in parallel as we did for sums and substractions, but their implementation would need special care. In fact, an addition 
 between two, $N$-bit integers requires $\sim N$ operations, while a simple longhand multiplication would be $\sim N^2$. While there exist efficient algorithm such 
 as Kabatsuba's \cite{kabatsuba1962multiplication}, which is $\sim N^{\log_2 3} = N^{1.58 \cdots}$, a detailed analysis would be required to assess whether they can
  be implemented in parallel,  and still provide a computational gain with respect to the serial case. 

\section*{Acknowledgments}

This work was granted access to the HPC resources of MesoPSL financed by the Region \^Ile de France. 

\appendix
\section{Serial update of the propensity functions}\label{app-1}

By using the definition  \eqref{equation-as} of the propensity functions,  the update rule \eqref{eq-update-occupancies} for the occupancies  and the definition \eqref{eq_nu}, we obtain the update rule for the propensity functions for each reaction: 
\begin{itemize}
	\item {$r=1$}: 
\begin{equation}\label{equation-a-nobit-1}
	\begin{aligned}
a_1 \ra & (\na -  1 ) (\nl + 1 ) = a_1 - \nl + \na - 1,\\
a_2 \ra &  (\na -1 ) \nr = a_2 - \nr,\\
a_3 \ra & (\nl +1 ) \nr = a_3 + \nr,\\
a_4 \ra & (\nl+1)\nl/2 = a_4 + \nl,\\
a_6 \ra & (\na-2)(\na-1)/2  = a_6 - \na +1,
	\end{aligned}
\end{equation}

\item {$r=2$}: 
\begin{equation}\label{equation-a-nobit-2}
	\begin{aligned}
a_1 \ra & (\na -1 ) \nl  = a_1 - \nl ,\\
a_2 \ra &  (\na -1 ) (\nr+1) = a_2 - \nr + \na -1 ,\\
a_3 \ra & \nl (\nr+1) = a_3 + \nl,\\
a_5 \ra & (\nr + 1)\nr/2 = a_5 + \nr ,\\
a_6 \ra & (\na-2)(\na-1)/2  = a_6 - \na +1,
	\end{aligned}
\end{equation}

\item {$r=3$}: 
\begin{equation}\label{equation-a-nobit-3}
	\begin{aligned}
a_1 \ra & (\na +2) (\nl - 1) = a_1 - \na + 2 \nl - 2,\\
a_2 \ra &  (\na + 2 ) (\nr - 1) = a_2 + 2 \nr - \na - 2 ,\\
a_3 \ra & (\nl -1)(\nr - 1) = a_3 - \nl - \nr + 1,\\
a_4 \ra & (\nl -1)(\nl-2)/2 = a_4 -\nl+1,\\
a_5 \ra & (\nr -1)(\nr-2)/2 = a_5 -\nr+1,\\
a_6 \ra & (\na +2)(\na+1)/2 = a_6 + 2\na+1.
	\end{aligned}
\end{equation}
\end{itemize}

In \cref{equation-a-nobit-1,equation-a-nobit-2}, the propensity functions which do not appear are not updated because they do not change.

\section{Parallel update of the propensity functions}\label{app-2}

Proceeding along the lines of \cref{app-1}, the update of the propensity functions in the parallel case reads:

\begin{itemize}
	\item {$r=1$}:
\begin{equation}\label{equation-a-1}
	\begin{aligned}
\bma_1 \ra & (\bna -  \changer ) (\bnl + \changer ) = \bma_1 - \bnlc + \bnac - \changer,\\
\bma_2 \ra &  (\bna - \changer ) \bnr = \bma_2 - \bnrc,\\
\bma_3 \ra & (\bnl + \changer ) \bnr = \bma_3 + \bnrc,\\
\bma_4 \ra & (\bnl+\changer)\bnl/2 = \bma_4 + \bnlc,\\
\bma_6 \ra & (\bna-2 \changer)(\bna-\changer)/2  = \bma_6 - \bnac + \changer
	\end{aligned}
\end{equation}

\item {$r=2$}: 

\begin{equation}\label{equation-a-2}
	\begin{aligned}
\bma_1 \ra & (\bna - \changer) \bnl  = \bma_1 - \bnlc ,\\
\bma_2 \ra &  (\bna - \changer) (\bnr+ \changer) = \bma_2 - \bnrc + \bnac - \changer ,\\
\bma_3 \ra & \bnl (\bnr+ 2 \changer ) = \bma_3 + \bnlc,\\
\bma_5 \ra & (\bnr + \changer)\bnr/2 = \bma_5 + \bnrc ,\\
\bma_6 \ra & (\bna-2 \changer)(\bna-\changer)/2  = \bma_6 - \bnac +\changer,
	\end{aligned}
\end{equation}

\item {$r=3$}: 
\begin{equation}\label{equation-a-3}
	\begin{aligned}
\bma_1 \ra & (\bna + 2 \changer) (\bnl - \changer) = \bma_1 - \bnac + 2 \bnlc - 2 \changer,\\
\bma_2 \ra &  (\bna + 2 \changer) (\bnr - \changer) = \bma_2 + 2 \bnrc - \bnac - 2 \changer ,\\
\bma_3 \ra & (\bnl - \changer)(\bnr -  \changer) = \bma_3 - \bnlc - \bnrc + \changer,\\
\bma_4 \ra & (\nl -\changer)(\nl-2 \changer)/2 = \bma_4 -\nl^\changer+\changer,\\
\bma_5 \ra & (\nr - \changer)(\nr-2 \changer)/2 = \bma_5 -\nr^\changer+\changer,\\
\bma_6 \ra & (\na +2 \changer)(\na+ \changer)/2 = \bma_6 + 2\na^\changer+\changer
	\end{aligned}
\end{equation}

\end{itemize}

Note that in \cref{equation-a-1,equation-a-2,equation-a-3} we rewrote the update rules in a computationally efficient way which uses sums, subtractions and multiplications by two only, but not multiplications between two arbitrary occupancy numbers, which would be computationally costly---see \cref{section-discussion}. Importantly, the update rules in \cref{equation-a-1,equation-a-2,equation-a-3} do not yield any error propagation as the update is iterated multiple times, because the integers are represented exactly as collections of zeros and ones.

\end{document}